\documentclass[sigconf]{acmart}
\acmConference[]{}{}{}

\author{Shihao Xia}
\affiliation{%
  \institution{Pennsylvania State University}
  \country{USA}
}

\author{Shuai Shao}
\affiliation{%
  \institution{University of Connecticut}
  \country{USA}
}

\author{Mengting He}
\affiliation{%
  \institution{Pennsylvania State University}
  \country{USA}
}

\author{Tingting Yu}
\affiliation{%
  \institution{University of Connecticut}
  \country{USA}
}

\author{Linhai Song}
\affiliation{%
  \institution{Pennsylvania State University}
  \country{USA}
}

\author{Yiying Zhang}
\affiliation{%
  \institution{University of California, San Diego}
  \country{USA}
}

\setcopyright{none}
\renewcommand\footnotetextcopyrightpermission[1]{}
\AtBeginDocument{%
  \providecommand\BibTeX{{%
    \normalfont B\kern-0.5em{\scshape i\kern-0.25em b}\kern-0.8em\TeX}}}

\setcopyright{acmcopyright}
\copyrightyear{2018}
\acmYear{2018}
\acmDOI{XXXXXXX.XXXXXXX}

\acmPrice{15.00}
\acmISBN{978-1-4503-XXXX-X/18/06}

\usepackage{graphicx} % Required for inserting images
\usepackage{enumitem}
\usepackage{tightenum}
\usepackage{listings}
\usepackage{tabularx}
\usepackage{array} 
\usepackage{diagbox}
\usepackage{setspace}
\usepackage{epsfig}
\usepackage{graphicx}
\usepackage{amsmath}
\usepackage[most]{tcolorbox}

\usepackage{url}
\usepackage{subcaption}
\usepackage[ruled]{algorithm2e}
\usepackage{booktabs}
\usepackage{multirow}
\usepackage{diagbox}

\newcommand{\shihao}[1]{{\color{orange} \sf (SH: #1)}}

\newcommand{\ignore}[1]{}

\newcommand{\boldunderpara}[1]{\noindent{\underline{\textbf{#1}}}}

\newcommand{\bolditalicparagraph}[1]{\noindent{{\textit{\textbf{#1}}}}}

\newcommand{\boldparagraph}[1]{\vspace*{1ex}\noindent{\textbf{#1}}}
\newcommand{\italicparagraph}[1]{\noindent\underline{\textit{#1}}}

\newcommand{\boldunderparagraph}[1]{\noindent\underline{\textbf{#1}}}

\newcommand{\Tool}{AuditGPT\xspace}

\newcommand{\beforecaption}{\vspace{-.15cm}\begin{spacing}{0.85}}
\newcommand{\aftercaption}{\vspace{-.15cm}\end{spacing}}
\newcommand{\mycaption}[3]{\beforecaption\caption{\label{#1}{#2} #3}\aftercaption}
\newcommand{\eg}{\textit{e.g.}}
\newcommand{\ie}{\textit{i.e.}}

\newcounter{insight}

\newcolumntype{Y}{>{\centering\arraybackslash}X}

\title{\Tool{}: Auditing Smart Contracts with ChatGPT}
\date{June 2023}

\begin{document}

\lstdefinelanguage{Solidity}{
  basicstyle=\ttfamily,
  stepnumber=1,
  numbersep=5pt,
  breaklines=true,
  showstringspaces=false,
  frame=single,
  tabsize=2,
  keywords=[1]{anonymous, assembly, assert, balance, break, call, callcode, case, catch, class, constant, continue, constructor, contract, debugger, default, delegatecall, delete, do, else, emit, event, experimental, export, external, false, finally, for, function, gas, if, implements, import, in, indexed, instanceof, interface, internal, is, length, library, log0, log1, log2, log3, log4, memory, modifier, new, payable, pragma, private, protected, public, pure, push, require, return, returns, revert, selfdestruct, send, solidity, storage, struct, suicide, super, switch, then, this, throw, transfer, true, try, typeof, using, value, view, while, with, addmod, ecrecover, keccak256, mulmod, ripemd160, sha256, sha3}, % generic keywords including crypto operations
	keywordstyle=[1]\color{blue}\bfseries,
	keywords=[2]{address, bool, byte, bytes, bytes1, bytes2, bytes3, bytes4, bytes5, bytes6, bytes7, bytes8, bytes9, bytes10, bytes11, bytes12, bytes13, bytes14, bytes15, bytes16, bytes17, bytes18, bytes19, bytes20, bytes21, bytes22, bytes23, bytes24, bytes25, bytes26, bytes27, bytes28, bytes29, bytes30, bytes31, bytes32, enum, int, int8, int16, int24, int32, int40, int48, int56, int64, int72, int80, int88, int96, int104, int112, int120, int128, int136, int144, int152, int160, int168, int176, int184, int192, int200, int208, int216, int224, int232, int240, int248, int256, mapping, string, uint, uint8, uint16, uint24, uint32, uint40, uint48, uint56, uint64, uint72, uint80, uint88, uint96, uint104, uint112, uint120, uint128, uint136, uint144, uint152, uint160, uint168, uint176, uint184, uint192, uint200, uint208, uint216, uint224, uint232, uint240, uint248, uint256, var, void, ether, finney, szabo, wei, days, hours, minutes, seconds, weeks, years},	% types; money and time units
	keywordstyle=[2]\color{teal}\bfseries,
	keywords=[3]{block, blockhash, coinbase, difficulty, gaslimit, number, timestamp, msg, data, gas, sender, sig, value, now, tx, gasprice, origin},	% environment variables
	keywordstyle=[3]\color{violet}\bfseries,
	identifierstyle=\color{black},
	sensitive=true,
	comment=[l]{//},
	morecomment=[s]{/*}{*/},
	commentstyle=\color{gray}\ttfamily,
	stringstyle=\color{red}\ttfamily,
	morestring=[b]',
	morestring=[b]",
  morecomment=[f][\color{red}]{-\ },
}

\lstdefinelanguage{GPTPrompt}{
basicstyle=\small,
  numbersep=5pt,
  breaklines=true,
  showstringspaces=false,
  breakindent=1em,
  frame=single,
}

\lstdefinelanguage{YAML}{
basicstyle=\small,
  numbersep=5pt,
  breaklines=true,
  showstringspaces=false,
  breakindent=1em,
  frame=single,
    keywords={true,false,null,y,n},
    keywordstyle=\color{darkgray}\bfseries,
    basicstyle=\ttfamily\footnotesize,
    sensitive=false,
    comment=[l]{\#},
    morecomment=[s]{/*}{*/},
    commentstyle=\color{purple}\ttfamily,
    stringstyle=\color{red}\ttfamily,
    morestring=[b]',
    morestring=[b]"
}

\begin{abstract}
%Today's smart contracts largely run on Ethereum, a decentralized blockchain platform. 
To govern smart contracts running on Ethereum, 
multiple Ethereum Request for Comment (ERC) standards have been developed, 
each containing a set of rules to guide the behaviors of smart contracts. 
Violating the ERC rules could cause serious security issues and financial loss, 
signifying the importance of verifying smart contracts follow ERCs. 
Today's practices of such verification are to either manually audit each single contract 
or use expert-developed, limited-scope program-analysis tools, both of which are far from being effective in identifying ERC rule violations.

This paper presents a tool named \emph{\Tool{}} that leverages large language models (LLMs) to automatically and comprehensively
verify ERC rules against smart contracts. 
To build \Tool{}, we first conduct an empirical study on 
222 ERC rules specified in four popular ERCs to understand their content, their security impacts, their specification in natural language, and their implementation in Solidity.
Guided by the study, we construct \Tool{} by separating the large, complex
auditing process into small, manageable tasks and design
prompts specialized for each ERC rule type to enhance LLMs’ auditing performance.
In the evaluation, \Tool{} successfully pinpoints 418 ERC rule violations  
and only reports 18 false positives, showcasing its effectiveness and accuracy.
Moreover, \Tool{} beats an auditing service provided by security experts in effectiveness, accuracy, and cost, demonstrating its advancement over state-of-the-art smart-contract auditing practices.

\end{abstract}

\maketitle

\section{Introduction}
\label{sec:intro}

\boldunderpara{Ethereum and ERC.}
%\songlh{The background of Ethereum and Solidity.}
Since the creation of Bitcoin, blockchain technology has evolved significantly.
One of the most important developments is Ethereum~\cite{eth-1,eth-2}, a decentralized, open-source blockchain platform. Ethereum enables the creation and execution of decentralized applications (DApps) like financial services and smart contracts, which are self-executing agreements with the terms of the contract directly written into code~\cite{dapps, sc-anatomy}. 

%\songlh{Briefly discuss what are ERCs and why there are rules. }
To govern smart contracts running on Ethereum, a set of formal standards called Request for Comments (\textit{ERCs}) have been developed~\cite{token-standard}. For example, the ERC20 standard defines a common set of rules for fungible tokens---tokens (digital assets) that are interchangeable with one another~\cite{erc20}. 
ERCs play a crucial role in the Ethereum ecosystem by providing a 
common set of rules and specifications that developers can follow when implementing smart contracts. ERCs ensure interoperability 
and compatibility between different Ethereum-based projects, 
wallets, and DApps~\cite{dapps}. 

\boldunderpara{ERC violation.}
%\songlh{Issues when ERC rules are violated}
The violation of an ERC could result in interoperability issues where a violating contract may not work properly with wallets or DApps. ERC violations could also lead to security vulnerabilities and financial loss. Additionally, ERC violations could result in de-listing of tokens from exchanges, as many exchanges have listing requirements of following ERC standards~\cite{erc-standard}.
Figure~\ref{fig:20-high} shows a violation of an ERC20 rule in a real smart contract. 
The \texttt{\_balances} field in line 2 monitors the number of tokens held by each 
address. Function \texttt{transferFrom()} in lines 6--10 facilitates the transfer of 
\texttt{amount} tokens from one address to another. 
%, as evidenced by the reduction \texttt{from}’s value in \texttt{\_balances} in line 14 and the increase of \texttt{to}’s value of in \texttt{\_balances} in line 15. 
%After the contract is deployed to 
%Ethereum, \texttt{transferFrom()} can be invoked by anyone through a message call.
ERC20 imposes multiple rules on \texttt{transferFrom()}, such as the necessity to fire an \texttt{Transfer} event for logging purposes and to
treat the transfer of zero tokens the same as transferring other amounts, both of which the contract follows. % to ensure consistent contract behavior. 
%Apparently, the contract adheres to both rules. 
However, the function \texttt{transferFrom()} violates a crucial ERC20 rule that mandates the function to verify whether the caller has the privilege to execute the transfer of \texttt{amount} tokens, which ensures financial security. 
Due to this violation, anyone can steal tokens from any address by invoking \texttt{transferFrom()} to transfer tokens to his address. 
The patch in line 7 illustrates how to fix the violation. 
The patch uses a two-dimensional map, \texttt{\_allowances}, to track
how many tokens ``from'' allows ``msg.sender'' to manipulate. 
The subtraction operation in this line triggers an exception and the termination of the transaction in case of underflow, thus preventing a caller of the function from transferring tokens if they do not have enough privilege.
%Thus, line 7 throws an exception and terminates the transaction when 
%the message caller does not have enough privilege to transfer \texttt{num} tokens, 
%satisfying the rule and preventing hackers from stealing tokens. 

{
\begin{figure}[t]

\begin{minipage}{\columnwidth}
\begin{center}
\scriptsize
\begin{lstlisting}[numbers=left,framexleftmargin=.15in,xleftmargin=.15in,language=Solidity,basicstyle=\ttfamily,morekeywords={-},morekeywords={+},keepspaces=true]
 contract ERC20 {
   mapping(address => uint256) _balances;
   mapping(address => mapping(address => uint256)) _allowances;
   event Transfer(address indexed _from, address indexed _to, uint256 _value);
   
   function transferFrom(address from, address to, uint256 amount) public returns (bool) {
+    _allowances[from][msg.sender] -= amount;
     _transfer(from, to, amount);
     return true;
   }
   function _transfer(address from, address to, uint256 amount) internal {
     require(from != address(0), "transfer from address zero");
     require(to != address(0), "transfer to address zero");
     _balances[from] -= amount;
     _balances[to] += amount;
     emit Transfer(from, to, amount);
   }
 }
\end{lstlisting}
\vspace{-0.05in}
\mycaption{fig:20-high}{An ERC20 rule violation that can be exploited to steal tokens.}
{\textit{(Code simplified for illustration purpose.)}}
\end{center}
\end{minipage}
\vspace{-0.35in}
\end{figure}
}

\boldunderpara{State of the art.}
Despite the importance of following ERC rules, it is hard for developers to do so, as they need a comprehensive understanding of all ERC requirements and their contract code. 
ERC standards include a huge number of rules. For just the four ERC standards that we study, there are 222 rules involved. 
%\yiying{good to get a concrete number if it it's just a range or something like 100s/1000s}\shihao{(e.g., four ERCs contain 222 rules in total)} \yiying{how many standards there are in total?}, some of which \yiying{talk about how hard it is to completely understand some rules like what what}
%This example underscores the challenges inherent in verifying the alignment 
%between contract implementations and ERC requirements. Programmers need a 
%comprehensive understanding of both the ERC requirements and the contract code 
%for effective verification. However, the number of rules can be substantial 
A simple operation could involve multiple rules.
For example, the \texttt{transferFrom()} function in Figure~\ref{fig:20-high} involves six rules from ERC20, %\yiying{from what standards?}\shihao{I guess it means purposes, like logging, functionality, and privilege check in table 1}, 
one of which was overlooked by the programmers causing the issue described above.
To make matters worse, these rules are described in different manners, with some being code declarations and others being natural languages. 

Meanwhile, contract implementations are often complex as well. 
One contract and its dependent code
usually contain hundreds to thousands of lines of source code in multiple files.
%\yiying{can you guys get some stats on how many lines of code open-source contracs each has?}
%\shihao{On average(200 contracts from our large dataset), each contract file contains 847.7 lines of Solidity source code.}
Some code details may be obscured within intricate 
caller-callee relationships, while others may involve numerous 
objects and functionalities possibly written by different 
programmers. All these complexities in ERC rules and smart 
contracts make it extremely hard for programmers to manually 
check for ERC violations. As a result,
ERC rule violations widely exist in real-world smart contracts~\cite{humanaudited}.

%\commentty{to be completed.}
%, making it hard for programmers . Consequently, programmers may 
%easily overlook certain code details, leading to rule violations.

Today's practices to avoid ERC violations are on two fronts. 
First, to avoid programmers' manual checking efforts, there have been efforts to develop program-analysis tools to automatically verify certain criteria of smart contracts~\cite{slither-erc,erc20-verifier}. 
These tools are limited by the types of verification they support. They only verify whether interfaces in smart contracts meet basic requirements, such as the declaration of all functions required by the corresponding ERC. %\yiying{required by what?}
These tools cannot detect complex requirements like the violated rule in Figure~\ref{fig:20-high}.
A fundamental reason why these and potentially future program-analysis tools cannot cover a wide range of verifications is that many 
ERC rules involve semantic information and require customization for individual contracts---a process that is time-consuming if 
not unfeasible. 
Second, there are several auditing services provided by security experts~\cite{certik,revoluzion,pixelplex,blockhunters, immunebytes, antier,humanaudited}.
Although auditing services are more thorough and comprehensive than program-analysis tools, 
they usually involve vast amounts of manual auditing efforts behind the scenes. Thus, these services are often costly and involve lengthy auditing periods.
Overall, auditing services are only suitable for limited users and use cases.
%unsuitable for the vast number of new contracts deployed on Ethereum each day.

\vspace{0.05in}
\noindent \textit{Are there any cheaper, automated, and thorough ERC rule verification methods?}
\vspace{0.05in}

\boldunderpara{Our proposal.}
This paper answers this research question in the affirmative by building \emph{\Tool}, 
a tool that leverages large language models (LLMs) to automatically audit smart contracts for ERC compliance. LLMs like ChatGPT have seen tremendous success in recent years and have %demonstrated their capabilities in many areas including natural language understanding, language translation, questions-and-answers, content generation, etc. LLMs have also 
been used in programming-language-related tasks like bug finding~\cite{sun2023gpt}, malware detection~\cite{csahin2021malware}, 
and program repairing~\cite{xia2023keep, ibbaleveraging}. %\yiying{fill these references and probably more PL-related use of LLM.}
However, as far as we know, there has been no exploration of LLMs in rule auditing, such as ERC compliance auditing.
%\commentty{or general rule auditing?} \shihao{Yes, I cannot find other paper for using LLM to do general rule auditing} 
The use of LLMs in ERC auditing introduces several new challenges.
First, there are many ERC rules with different natures and specified in different natural languages. 
It is unclear how to represent them effectively to an LLM.
Second, most ERC rules are specific to program semantics. Thus, we need a good representation of smart-contract code to an LLM.
Third, LLMs largely work as black boxes during the auditing process, and their
effectiveness largely depends on the prompts employed.
%How can we devise appropriate prompts to enhance the effectiveness of LLMs?

%\yiying{above correct? any other challenges?} \shihao{Another challenge is automating the auditing process?}
%\commentty{are there any existing LLM-based approaches for rule auditing other than ERC? Why is ERC so unique, or can our approach be generalized to handle other kinds of rules?} \shihao{NO? ERC is so unique because Ethereum is the most popular platform for smart contracts and ERCs are the standards for them. Other kinds of rules may include protocols or standards, like TCP, UDP, JVM, etc? or even simply the rules in the function(other programming languages) comment.}

To confront these challenges, we first conduct an empirical study of ERC standards to gain insights into ERC rules. Specifically, we examine four popular ERC standards and the 222 rules they encompass.
Our study aims to understand what the rules are about, the security impacts of violating them, and how they are articulated in natural language within the ERCs and implemented in smart contracts. Our study yields five insights that can benefit Solidity programmers, security experts, and ERC protocol designers. For example, we observe that approximately one-fifth of ERC rules focus on verifying whether an operator, token owner, or token recipient possesses adequate privileges to execute a specific operation. Violating these rules can result in a clear attack path, leading to financial loss (\eg, Figure~\ref{fig:20-high}).
Furthermore, we discover that most ERC rules can be validated within a single function. Additionally, we identify linguistic patterns used to express rules in ERCs and notice a correlation between the natural language specification of a rule and its implementation.

Second, we design and implement \Tool{} based on three principles.
\textit{Divide and conquer}: We split the auditing process into a startup 
phase for automatically extracting rules from ERCs and a working phase for inspecting individual contracts. 
Additionally, we divide large contracts into small code segments 
and instruct LLMs to inspect each segment individually against a 
specific rule to focus their attention.
\textit{Guided by our study}: We separate each individual contract based 
on its public functions. We leverage identified linguistic 
patterns as one-shot examples during ERC rule extraction.
\textit{Specialization}: We create specialized questions for each type of 
rules and automate the process from ERC rule extraction to 
generating specialized prompts using our designed informative 
YAML format. Furthermore, we craft one-shot Solidity code 
examples for specific rules to enhance LLMs' auditing effectiveness.
The mechanisms employed in \Tool{} can serve as inspiration for 
future researchers exploring the application of LLMs in other 
programming-language-related tasks.

We construct two datasets to assess \Tool{}: a large dataset comprising 200 contracts randomly selected from etherscan.io~\cite{etherscan} and polygonscan.com~\cite{polygonscan}, and another dataset consisting of 30 contracts with ground-truth violation labeling (performed by us) and human-auditing reports crafted by security experts at the Ethereum Commonwealth Security Department (ECSD)~\cite{humanaudited}. 
\Tool{} identifies 279 ERC rule violations from contracts in the large dataset, with four violations exhibiting a clear attack path leading to potential financial losses (one is shown in Figrue~\ref{fig:20-high}). Additionally, \Tool{} reports only 15 false positives. Overall, \Tool{} is effective and accurate in auditing smart contracts and identifying ERC rule violations. 
We compare \Tool{} with an automated program analysis technique~\cite{slither-erc} and the human auditing service~\cite{humanaudited} with the small dataset. 
\Tool{} detects 50\% more 
violations than the two baseline solutions. 
Furthermore, \Tool{} reduces time and monetary costs by a thousand fold compared with the human auditing service. 
%Overall, \Tool{} propels the state-of-the-art in contract auditing.

%We have made \Tool{} and all our study results publicly available ~\cite{XXX}. 
In sum, we make the following contributions. 

\begin{itemize}[leftmargin=3em]
    \item We conduct the first empirical study on Ethereum ERC rules made for smart-contract implementations.
    
    \item We design and implement \Tool to audit smart contracts and pinpoint ERC rule violations. 
    
    \item We conduct thorough experiments to assess \Tool{} and confirm its effectiveness, accuracy, and advancement. 

\end{itemize}

%DApps running on Ethereum are most commonly written in Solidity~\cite{soliditylang}, a high-level programming language specifically designed for writing smart contracts on the Ethereum blockchain. \yiying{I feel that Solidity is not core to this work. no need to talk about it here.}

\section{Background}
This section gives the background of the project, including 
Solidity smart contracts, 
ERCs, 
and existing techniques related to ours.

\subsection{Ethereum and Solidity Smart Contracts}

Ethereum is a blockchain system that enables programmers to 
create and deploy smart contracts for the development 
of decentralized applications~\cite{eth-1,eth-2}. 
Both Ethereum users and smart contracts are represented by 
distinct Ethereum addresses, which can be used to send and receive Ethers 
(the native Ethereum cryptocurrency) and interact with smart contracts, 
thereby leveraging their functionalities for conducting complex transactions. 
Ethereum cultivates a thriving digital economy ecosystem. 
At the time of writing, the price of one Ether is over \$2K, 
with the total market value of all Ethers exceeding \$200B~\cite{eth-price}. 
%ranking second only to Bitcoin among all cryptocurrencies~\cite{eth-price}. 
Daily transactions on Ethereum surpass one million, 
with a volume exceeding \$4B~\cite{eth-daily}. 
Smart contracts play a pivotal role in Ethereum's success, as they guide 
the majority of transactions and enable crucial 
functionalities~\cite{erc20,erc721,eth-defi} 

%such as cryptocurrencies~\cite{erc20}, NFTs~\cite{erc721}, and decentralized finance~\cite{eth-defi}.

Solidity stands out as the most widely used programming language
for writing smart contracts~\cite{solidity-popular-1,solidity-popular-2}. 
With a syntax resembling ECMAScript~\cite{ecmascript}, Solidity effectively conceals 
the intricacies of the Ethereum blockchain system. 
Implementing a contract in Solidity is similar to implementing 
a class in Java. A contract contains 
contract fields (state variables) to store the contract's states 
and functions to realize its functionalities. 
%Analogous to object-oriented programming languages, 
A function in Solidity can be public, internal, or private. 
Public functions serve as the contract's interface, providing external access 
to its functionalities. These functions can be invoked by a different contract 
or an Ethereum user through a message call, 
while private or internal functions cannot. 
Additionally, contracts can define events emitted during execution, 
serving as logs on-chain 
that can be analyzed by off-chain applications. 

An example of a contract is presented 
in Figure~\ref{fig:20-high}. The contract has two contract fields in lines 2 and 3, 
%\texttt{\_balances} (line 2) and \texttt{\_allowances} (line 3), 
%tracking the number of tokens owned by each address and 
%the tokens approved by the first dimension for manipulation 
%by the second dimension, respectively. 
and defines event \texttt{Transfer()} 
in line 4 and emits it in line 16. The public function \texttt{transferFrom()} 
(lines 6--10) can be called by any Ethereum user or contract after the contract is deployed, while the internal function \texttt{\_transfer()} (lines 11--17) is restricted to calls from the same address.

\subsection{Ethereum Request for Comment (ERC)}

ERCs serve as technical documents that outline the requirements 
for implementing smart contracts, ensuring interoperability and 
compatibility across various contracts, applications, and platforms 
and helping foster the Ethereum ecosystem~\cite{erc-eip1, erc-standard, stefanovic2023proposal}. 
Typically, an ERC begins with a concise motivation. For example, ERC20 emphasizes its
role in defining a standard token interface for tokens to be manipulated by 
applications like wallets and decentralized exchanges~\cite{erc20}.
Subsequently, an ERC details all necessary public functions and events, specifying their parameters, return values, and optional attributes for the parameters. 
%Functions and events play distinct roles: functions are invoked 
%to interact with contracts, while events are used to trace crucial contract actions. 
Additionally, an ERC typically outlines requirements through plain text  or code comments for each function or event 
around its declaration. 
For instance, %ERC20 defines nine functions and two events. 
beyond the requirements for the function API and return value generation, 
ERC20 incorporates four additional rules for \texttt{transferFrom()} 
(\eg, Figure~\ref{fig:20-high}): 
mandating the emission of a \texttt{Transfer} event, 
verifying whether the message sender is approved to manipulate 
the token owner’s tokens and throwing an exception if not, 
treating the transfer of zero tokens similarly to other amounts, 
and emitting an event when transferring zero tokens.

Violating ERC rules can result in substantial financial losses and unexpected contract behaviors. For instance, ERC721 mandates \texttt{onERC721Received()} to be called for each token transfer, when the recipient is a contract. Additionally, it further requires that 
the caller must verify that the return is a specific magic number. 
These two rules ensure that the recipient contract has 
the capability to handle the transferred tokens. 
Transferring tokens to a contract lacking this capability can result 
in the tokens being permanently trapped in the recipient contract. 
This issue was initially reported in 2017 on Ethereum Reddit, 
leading to a loss of \$10,000 worth of tokens at that time, and has since caused millions of dollars in losses~\cite{erc20-problem-history}.
As another example, the rule violation in Figure~\ref{fig:20-high} 
%fails to adhere to the rule of verifying the privilege of its message caller, and thus 
opens the door for a hacker to
pilfer tokens from any account. 
In summary, it is crucial to ensure that contracts adhere to ERC rules to safeguard financial assets and 
ensure the proper functionality of the contracts.

%\songlh{An example of a code implement ERC-20}

%\mengting{The ERC20 Token is the minimum standard template for fungible tokens. It defines 6 mandatory functions and 2 events. 3 of these functions are getter function of state variables which represent account balance (line 2 in Figure~\ref{fig:20-high}), total token supply (line 3) and the amount one address can spend on behalf of another address (line 4). Except for the interface, the standard also includes rules about implementing functions and rules about emitting the events. For example, for the function transferFrom which transfers '\_value' amount of token from address '\_from" to address '\_to', the Transfer event must be fired as shown in line 19 in Figure~\ref{fig:20-high}. This function should throw unless the \_from account has deliberately authorized the sender of the message via some mechanism. For example, as shown in line 10 in Figure~\ref{fig:20-high}, if amount is larger than \_allowances[from][msg.sender] which means the amount it wants to transfer exceeds the amount the address 'from' allows the message sender to transfer, this line will throw an exception because of underflow. Moreover, transfers of 0 values must be treated as normal transfers. As shown in line 10-19 in Figure~\ref{fig:20-high}, the function transferFrom does not have special handling for zero value of amount. In addition, it also defines 3 optional functions which return the name, symbol and number of decimal places of the token. }

\subsection{Related work}
\label{sec:related}

There are tools designed to automatically pinpoint ERC rule violations. 
Slither contains specific checkers (\ie, slither-check-erc~\cite{slither-erc}) that
scrutinize whether a given contract adheres to the corresponding ERC 
requirements for 11 ERCs. 
However, these checkers have limited functionalities; they primarily focus on 
verifying the presence of required functions and events, 
ensuring that the declarations of these functions and events 
align with the specified requirements, and confirming that functions emit 
the necessary events. Unfortunately, these checkers lack the capability 
to inspect more advanced requirements, such as determining if the 
message sender has the necessary privilege to transfer a token,
and thus, they miss complex violations, that can be detected by \Tool{} (see Section~\ref{sec:compare}). 
ERC20 verifier is a tool dedicated to ERC20 contracts~\cite{erc20-verifier}.
%It specifically examines whether the declarations in the implementation match the ERC20 declaration requirements and whether functions 
%correctly emit events. 
Its assessments are 
similar to the checks performed by Slither.

Researchers have developed automated tools for identifying various Solidity bugs, including reentrancy bugs~\cite{liu2018reguard, qian2020towards, xue2020cross}, nondeterministic payment bugs~\cite{wang2019detecting, li20safepay}, consensus bugs~\cite{yang2021finding, chen2023tyr}, eclipse attacks~\cite{wust2016ethereum, xu2020eclipsed, marcus2018low}, out-of-gas attacks~\cite{grech2018madmax,ghaleb2022etainter}, 
and code snippets consuming unnecessary gas~\cite{kong.pattern,brandstatter2020characterizing, brand.efficiency,nelaturu2021smart,chen2018towards,gaschecker,chen2017under,slither}. 
Unfortunately, these techniques focus on bugs that are unrelated to program semantics 
and may manifest in any Solidity or Ethereum contract. 
Consequently, they are unable to identify ERC rule violations like \Tool{} does, 
as rule violations stem from the failure to meet specific semantic 
requirements outlined by individual ERCs.

Security experts offer auditing services to identify security vulnerabilities 
or logic flaws in Solidity 
contracts~\cite{certik,revoluzion,pixelplex,blockhunters,immunebytes,antier,humanaudited}. Some of these services also assess ERC compliance. 
However, these auditing services come with considerably higher financial 
cost compared to automated tools, and Solidity programmers may be hesitant to 
utilize them. Furthermore, manual contract auditing is a time-consuming 
process and it is 
challenging for it to handle 
the amount of contracts deployed on Ethereum daily. 

Researchers have already utilized LLMs provided by ChatGPT for analyzing Solidity code,
such as detecting vulnerabilities~\cite{sun2023gpt}, patching bugs~\cite{ibbaleveraging},
and testing Solidity programs~\cite{alici2023openai}.
However, the problem we address differs from existing techniques. 
Given that ERCs specify a multitude of implementation rules and most of them are related to program semantics, 
we essentially resolve bugs in a much wider array of types than existing techniques.

%Researchers have already utilized LLMs provided by ChatGPT for analyzing Solidity code. 
%Sun \etal{} define features for ten vulnerability types and instruct ChatGPT to inspect Solidity contracts to detect vulnerabilities of those types~\cite{sun2023gpt}. 
%Ibba \etal{} explore leveraging ChatGPT for patching bugs in Solidity programs~\cite{ibbaleveraging}. 
%Alıcı \etal{} seek guidance from ChatGPT on testing a Solidity program, then follow the instructions provided by ChatGPT to execute the testing process with ChatGPT and identify bugs~\cite{alici2023openai}. 
%The problem we address differs from existing techniques. 
%Given that ERCs specify a multitude of implementation rules and most of them are related to program semantics, 
%we essentially resolve bugs in a much wider array of types than existing techniques.

%\mengting{
%CertiK~\cite{certik}, Revoluzion~\cite{revoluzion}, PixelPlex~\cite{pixelplex} and BLOCKHUNTERS~\cite{blockhunters} provides manual auditing service to assess smart contracts. ImmuneBytes~\cite{immunebytes} also provides smart contract auditing service to check the compliance, evaluate models and identify vulnerabilities. Antiersolutions~\cite{antier} audits the security of smart contracts, updates threat model and provides enterprise security counseling.}

\section{Empirical Study on ERC rules}
\label{sec:study}

This section presents our empirical study on implementation rules specified in the ERC documents, including the methodology employed for the study and the categories established for the rules. 
%and the findings associated with the study. 

\subsection{Methodology}

%\boldparagraph{Studied ERCs.}
From the 84 ERCs in the final status, we choose ERC20, ERC721, ERC1155, 
and ERC3525 as our study targets. We base our selection on several criteria, 
including their popularity with numerous corresponding contracts, 
their complexity with various requirements, 
and their significance in the Ethereum ecosystem.

\emph{ERC20} is a technical standard for fungible tokens (\eg, cryptocurrencies) and is probably the most famous ERC standard.
It outlines operational requirements for minting, burning, and transferring 
tokens~\cite{erc20}. 
Presently, there are over 450,000 ERC20 tokens on the Ethereum 
platform~\cite{erc20-popular}, with many boasting a market capitalization surpassing \$1 billion (\eg, USDT~\cite{USDT}, SHIB~\cite{SHIB}, Binance USD~\cite{Binance}).

\emph{ERC721} is designed for non-fungible tokens (NFTs), where each token is 
distinct and indivisible~\cite{erc721}. ERC721 specifies how ownership of NFTs 
is managed and 
stands as the most popular NFT standard. 
It is adhered to by major NFT marketplaces~\cite{opensea,rarible}.

\emph{ERC1155} aims to enable a single contract to oversee both fungible and 
non-fungible tokens~\cite{erc1155}. Additionally, it facilitates batch 
operations. For example, multiple types of tokens can be transferred from one 
address to another together. 
ERC1155 has found adoption in various gaming and 
charity donation projects~\cite{Horizon,9Lives,Reewardio}.

\emph{ERC3525} is tailored for semi-fungible tokens, where each token is distinct like NFTs but incorporates an additional qualitative nature like fungible tokens~\cite{erc3525}.
ERC3525 has already been leveraged for financial instruments~\cite{fujidao,bufferfinance}.

We carefully review the official documents (including text descriptions and associated code)
and manually identify rules by evaluating their relevance to contract implementations, 
whether they have clear restricting targets, and whether they offer actionable checking criteria. 
To maintain objectivity, all extracted rules are examined by at least two paper authors. 
Certain rules are extracted from the textual descriptions, 
with some explicitly using terms like ``must'' or ``should'' to convey obligations. 
The remaining rules are gained from code sections pertaining to function and event declarations.
%indicating the requirements for contract interfaces.
In total, we identified 222 rules from the four ERCs. 
%Table~\ref{tab:rules} illustrates the distribution of these rules across the ERCs.

Our study primarily answers three key questions regarding the identified rules: 
1) what rules are specified? 2) why are they specified? and 
3) how are they specified in the ERCs and implemented in contracts? 
The objective is to garner insights 
for building techniques to automatically detect rule violations. 
%Similar to the process of rule identification, 
All study results are carefully reviewed by at least two paper authors.

%\subsection{Study Results}

\subsection{Rule Content (What)}
We first inspect the content of the 222 ERC rules to understand what
checks we need to detect their violations. As shown by the rows in Table~\ref{tab:study},
we separate the ERC rules into four categories. 

\italicparagraph{Privilege Checks.} 
46 rules delineate the necessary privileges for executing specific token operations. 
Among them, 14 pertain to verifying if the operator (\eg, a message caller) possesses
the required privilege. For example, the implementation in Figure~\ref{fig:20-high}
violates an ERC20 rule about \texttt{transferFrom()}, 
which stipulates that the implementation
must verify the message caller is authorized by the token owner to
transfer the \texttt{amount} tokens.
Additionally, 13 rules address whether the token owner holds sufficient privilege.
For example, function \texttt{transfer(address to, uint256 value)()} in ERC20 
sends \texttt{value} tokens from the message caller (the token owner) to recipient \texttt{to}, 
and ERC20 mandates the message caller has enough tokens. 
%However, ERC20 does not have such a requirement for \texttt{transferFrom()}.
The remaining 19 rules focus on the token recipient when the operation is a transfer. 
For instance, ERC721 specifies the recipient cannot be address zero for \texttt{safeTransferFrom()}. 
Moreover,
it also mandates calling \texttt{onERC721Received()} on the recipient when transferring tokens to
a contract. 
It further requires the caller to check whether the return value of \texttt{onERC721Received()} is a magic number and mandates the caller to 
throw an exception if not. Both ERC1155 and ERC3525 have similar rules.

%and the \texttt{require} in line 11 ensures adherence to this rule.

% \shihao{The remaining 7 rules focus on the error handling of the transfer process. Different ERC has different way to handle the error during the transferring. For instance, \texttt{transferFrom()} in Figure~\ref{fig:20-high}, the boolean return indicates the success of the transfer. However, ERC1155 requires transfer-related functions must revert on any other error instead of returning false.}

\italicparagraph{Functionality Requirements.}
Among the 62 rules governing code implementation, 
43 rules specify how to generate the return value for a function. 
For example, ERC1155 mandates that \texttt{balanceOf(address \_owner, uint256 \_id)} should return the amount of tokens of type \texttt{\_id} 
owned by \texttt{\_owner}. 
In particular, when an ERC defines a %\shihao{transfer-related} 
function with a Boolean return, 
we interpret it as an implicit requirement for the function 
to return \texttt{true} upon successful completion and \texttt{false} otherwise. 
Ten rules focus on managing input parameters. For instance, ERC20 dictates that \texttt{transferFrom()} should treat 
scenarios where zero tokens are transferred the same way 
as cases when non-zero tokens are transferred. 
The implementation shown in Figure~\ref{fig:20-high} adheres to this rule. 
Four rules explicitly mandate the associated function to 
throw an exception when any error occurs or the recipient rejects a transfer. 
Four rules specify how to update particular variables. For example, one ERC1155 rule for transferring multiple tokens together  
requires the balance update for each input token type to 
follow their order in the input array. 
The last rule is mandated by ERC3525, which allows a contract to manage multiple types of fungible tokens, 
where tokens within the same slot belong to the same type. 
This rule requires that when transferring tokens, the slot of the recipient that receives the tokens must be the same as the slot the sender uses. 

\begin{table}[t]
\centering
\small

\mycaption{tab:study}
{ERC rules' content and security impacts.}
{
}
{

\begin{tabular}{|l|c|c|c||c|}
\hline
\diagbox{\textbf{content}}{\textbf{impact}}
                 &  {\textbf{High}} & {\textbf{Medium}} & {\textbf{Low}}  & {\textbf{Total}} \\

\hline
\hline

{\textbf{Privilege Check}}     & 46 &  0 &   0  &  46   \\ \hline
{\textbf{Functionality}}      & 22  & 40  &   0  & 62  \\ \hline
{\textbf{Usage}}   & 0 &  56 &   0  & 56 \\ \hline 
{\textbf{Logging}}   & 0 &  0 &  58 & 58 \\ \hline 
\hline
{\textbf{Total}}     & 68 &  96 &  58 & 222 \\ \hline

\end{tabular}

}

\end{table}

\italicparagraph{Code Usage.}
56 rules are about how a piece of code is used.
For example, the four ERCs mandate the declaration of 52 functions,
as users may interact with contracts following the interfaces 
outlines in the ERCs after the contracts are deployed. 
ERC20 specifies that three functions are optional and necessitates checks for their existence before invoking those functions. 
Additionally, ERC20 requires that if a function returns 
a Boolean value, the caller of the function must inspect the 
return and cannot assume the call always succeeds. 

%ERC721 mandates calling \texttt{onERC721Received()} on the recipient when conducting a token transfer. 
%It further requires the caller to check whether the return value of \texttt{onERC721Received()} is a magic number and mandates the caller to 
%throw an exception if not. Both ERC1155 and ERC3525 have similar rules.

\italicparagraph{Logging.}
ERCs impose logging requirements by emitting events. 
For example, ERC20 mandates the emission of a \texttt{Transfer} event 
when a transfer operation occurs (\eg, line 14 in Figure~\ref{fig:20-high}). 
In total, there are 58 rules related to logging. 
Out of these, 32 specify when an event should be emitted, 
while the remaining 26 pertain to event declarations, including 
15 rules specifically mandating attributes of event parameters.

\stepcounter{insight}
\boldunderparagraph{Insight \arabic{insight}:}
{\it{
Given that ERC rules primarily involve contract semantics, constructing 
program analysis techniques for identifying ERC rule violations across diverse 
contracts poses a significant challenge.
}}

We further study the valid scope for each rule. 
The valid scopes of 200 rules are confined to a single function. 
For instance, ERC20 mandates that \texttt{transferFrom()} 
in Figure~\ref{fig:20-high} scrutinizes whether the message caller 
possesses the privilege to handle the token owner's tokens. 
As another example, 43 rules concern how a function generates its return value. 
Moreover, 11 rules pertain to event declarations. 
In the remaining 11 cases, their valid scopes encompass the entire contract. 
For instance, ERC20 necessitates emitting a \texttt{Transfer} event 
whenever a token transfer occurs. Similarly, ERC721 requires calling \texttt{onERC721Received()} on the recipient contract for each token transfer, 
along with checking the return value of the function.

\stepcounter{insight}
\boldunderparagraph{Insight \arabic{insight}:}
%\noindent{\textbf{Insight \arabic{insight}:}}
{\it{
Most ERC rules can be checked within a limited scope (\eg, a function, an event declaration site), and there is no need to analyze the entire contract for compliance with these rules. 
}}

\subsection{Violation Impact (Why)}

We analyze the security implications of rule violations to understand the reasons behind the rule specifications. As shown by the columns Table~\ref{tab:study}, 
we categorize the rules’ impacts into three levels.

\italicparagraph{High.}
A rule is deemed to have a high-security impact if there exists a clear 
attack path exploiting its violation, 
resulting in financial loss. 
As shown in Table~\ref{tab:study}, 68 rules fall into this category, 
encompassing all rules related to privilege checks. 
For instance, the failure to verify an operator's privilege can enable a hacker to pilfer tokens (\eg, Figure~\ref{fig:20-high}) and neglect to inspect whether a recipient address
is non-zero can lead to tokens being lost permanently. 
Concerning rules associated with the implementation of specific functions, 
non-compliance with 22 of them can also result in financial loss. 
Among these, 18 rules outline how to generate return values 
representing token ownership, such as \texttt{balanceOf()} returning 
the number of tokens belonging to an address and \texttt{ownerOf()} designating the owner address of the input token. Errors that fail to provide accurate returns for 
these functions can lead to scenarios where a valid token is trapped 
in an address, or an address sends out tokens not belonging to it. 
The remaining four are related to properly updating tokens' ownership.
%
%updating privilege-related variables. ERC20 mentioned in \texttt{approve(spender, value)} that value should overwrite the current allowance. Breaking this rule, for example, by increasing the allowance, might accidentally grant the spender more funds than intended, potentially leading to a loss of balance.
%
%}
%
% mandated by ERC3525, 
For example, ERC3525 mandates that the receiver's slot 
of transferred tokens must match the sender's. 
As each slot in ERC3525 represents a type of tokens, 
violating this rule can cause tokens with a high value to 
transform into tokens with a low value after a transfer. 
ERC721, ERC1155, and ERC3525 all require inspection 
of whether an address has the capability to handle received tokens 
during a token transfer to prevent tokens from becoming 
trapped in the recipient address. These rules pertain to code usage and also have a high impact.

\italicparagraph{Medium.}
A rule is deemed to have a medium impact if its violation can lead 
to unexpected contract or transaction behavior, while not having a clear attack path 
to causing financial loss. For example, if a public function's API fails 
to adhere to its ERC declaration requirement, 
invoking the function with a message call following the requirement 
would trigger an exception. Another example is ERC20's requirement 
that the function \texttt{transferFrom()} (\eg, Figure~\ref{fig:20-high}) treats the 
transfer of zero tokens the same way as transferring non-zero values. 
If a contract does not adhere to this rule, its behavior would be 
unexpected for the message caller.

\italicparagraph{Low.}
All event-related rules are about logging. We consider their security impact as low.

\stepcounter{insight}
\boldunderparagraph{Insight \arabic{insight}:}
%\noindent{\textbf{Insight \arabic{insight}:}}
{\it{
For numerous rules, their violations present a clear attack path 
for potential financial loss, emphasizing the urgency of detecting and 
addressing these violations.
}}

\begin{table}[t]
    \centering
    \small
    \setlength{\tabcolsep}{3pt} 
    \mycaption{tab:linguistic}
    {Linguistic Patterns.}
    {\textit{([*]: a parameter in a linguistic pattern, and \{*\}: an optional parameter.
    P: privilege checks, F: functionality requirements, U: code usage, and L: logging.
    [subject] could be a function or an event. [must] could be ``must'', ``must not'', and ``should''.
    [action] could be ``handle'',  ``fire'', and ``throw''.
    [assign] could be ``be the'' and ``be set to''.
    [role] could be ``an authorized operator''.
    )}
    }
    \begin{tabular}{|l|l|c|c|c|c|c|}
    \hline
\multirow{2}{*}{{\textbf{ID}}} & \multirow{2}{*}{\textbf{Patterns}} & \multicolumn{4}{c|}{\textbf{Content}} & \multirow{2}{*}{{\textbf{Total}}}       \\ \cline{3-6}
 &  & {{\textbf{P}}} & {{\textbf{F}}} & {{\textbf{U}}}  &  {{\textbf{L}}} &  \\ \hline \hline
    
    CP1 & [subject] [must] [action] \{condition\}             & 17  & 7   & 4 & 0  & 28        \\ \hline
    CP2 & [action] [must] result in revert                    & 3   & 0  &  0 & 0   & 3        \\ \hline
    CP3 & Caller must be approved to [action]               & 2   &0    &0&  0   &  2       \\ \hline
    CP4 & [must] revert [condition]                           &  14  & 4& 0&  0   & 18       \\ \hline
    CP5 & Caller [must] be [role]                             &  2  &0&0&   0  &   2      \\ \hline
    CP6 & [action] is considered invalid                    &  0  &1&0&   0  &   1      \\ \hline

    CP7 & [condition] [subject] [must] call [function]        & 8   & 0 & 0 & 0     & 8    \\ \hline
    
    EP1 & [must] [action] [event] \{condition\}               & 0  &0&0&  6    &   6      \\ \hline
    EP2 & [event] emits \{condition\}                       & 0   &0&0&  25   &     25    \\ \hline
    EP3 & \{condition\} without emitting [event]            & 0  &0&0&    1  &  1       \\ \hline
    RP1 & return                                          & 0 & 15&0 & 0 &      15   \\ \hline
    RP2 & @return/@notice                                 &  0  & 28 & 0&   0  &    28     \\ \hline
    AP1 & [subject] [must] [assign]             & 0 & 7& 0 &  15     & 22        \\ \hline \hline

{\textbf{Total}}  &  & 46 &  62 &  4 & 47 & 159\\ \hline 
\end{tabular}
\end{table}

\subsection{Specification and Implementation (How)}

Among the 222 rules, 63 pertain to function or event declarations, 
and they are precisely outlined by providing the correct declaration using 
Solidity code. The remaining 159 rules are articulated 
in natural language texts. We have identified 12 linguistic patterns 
to categorize how these rules are presented. As shown in Table~\ref{tab:linguistic}, 
four patterns encompass more than 20 rules each. CP1 and RP2 are 
the most widely employed patterns, both covering 28 rules. 
CP1 is primarily utilized for conducting privilege checks. 
For instance, ERC20 specifies the violated rule in Figure~\ref{fig:20-high} as 
``the function SHOULD throw unless the \_from account 
has deliberately authorized the sender of the message via some mechanism.’’ 
RP2 is employed to delineate return values. 
Conversely, there are patterns covering a very limited number of rules; 
CP6 and EP3 are each associated with only one rule.

\stepcounter{insight}
\boldunderparagraph{Insight \arabic{insight}:}
%\noindent{\textbf{Insight \arabic{insight}:}}
{\it{
The majority of rules can be specified using common linguistic patterns, while 
there are a few rules that are specifically outlined in natural language in a distinct manner.
}}

We further categorize the 12 patterns into six groups based on how these rules are implemented in Solidity. 
Patterns with IDs sharing the same prefix are grouped together in Table~\ref{tab:linguistic}. 
Group CP encompasses seven linguistic patterns where rule implementations involve a condition check 
followed by the execution (or non-execution) of an action 
if the check passes. 
This condition check may be explicitly implemented using an \texttt{if} or a \texttt{require} statement, or it could be implemented implicitly. 
For instance, the check required by the violated rule in Figure~\ref{fig:20-high} can be performed using a subtraction operation, as illustrated by line 7. 
Group EP pertains to rules related to emitting or not emitting events, with implementations or violations involving the keyword \texttt{emit}. Similarly, group RP involves rules where implementations revolve around \texttt{return}, and group AP deals with rules where implementations involve updating field values.

\stepcounter{insight}
\boldunderparagraph{Insight \arabic{insight}:}
%\noindent{\textbf{Insight \arabic{insight}:}}
{\it{
How a rule should be implemented usually correlates with how the rule is specified in the ERC. 
}}

\section{\Tool{} Design}

\Tool{} utilizes an LLM to verify whether a contract implementation 
adheres to ERC requirements. 
Consequently, the design of \Tool{} primarily centers around 
creating appropriate LLM prompts, including how to specify ERC requirements 
in prompts, how to supply contract code to the LLM, 
and strategies to enhance the effectiveness and accuracy of violation detection.
This section commences with an overview of \Tool{}, 
followed by a detailed presentation of the three design aspects of prompt construction.

\subsection{Overview}

%As shown by Figure~\ref{fig:overflow}, 
\Tool{} takes a startup phase to analyze ERCs and a subsequent working phase dedicated to inspecting individual contracts.

We observe that an LLM is more effective when assessing the 
satisfaction of individual rules rather than concurrently checking multiple rules. 
Consequently, we adopt a rule sequentialization policy, 
wherein we iteratively guide the LLM through all rules in multiple prompts, 
checking only one rule in each prompt. 
To facilitate this process, during the startup phase, \Tool{} employs 
the LLM to extract rules from an ERC and express them in a specific YAML format to 
streamline prompt construction during the working phase. 
\Tool{} achieves this by first instructing the LLM to enumerate all public 
APIs of an ERC and subsequently requesting the LLM 
to list rules for each API. For example, \Tool{} identifies nine functions and 
two events for ERC20 and extracts five rules for 
the function \texttt{transferFrom()} in Figure~\ref{fig:20-high}.

Similarly, we notice \Tool{} exhibits better performance 
when scrutinizing smaller code segments separately
compared to assessing the entire contract implementation. 
Consequently, during the working phase, \Tool{} individually 
inspects each public function. For every public function, \Tool{} apply code 
slicing to compute 
its associated code and then employs individual prompts 
to inquire whether the public function, along with its related code, 
adheres to each rule extracted for the function during the startup phase. 
For instance, \Tool{} generates five prompts to 
ascertain whether function \texttt{transferFrom()} 
in Figure~\ref{fig:20-high} 
satisfies the five rules. 
Moreover, \Tool{} employs mechanisms like prompt specialization 
and one shot~\cite{duan2017one} to enhance its effectiveness. 
For example, after employing these mechanisms, an additional 4 prompts are generated for \texttt{transferFrom()} in Figure~\ref{fig:20-high}.

\subsection{ERC Rule Extraction}

\Tool{} follows a three-step process to extract rules from an ERC and stores them in a YAML file.

First, \Tool{} supplies the ERC to its LLM and 
instructs it to enumerate all functions in the ERC, presenting the information in a YAML array. 
Each element of the array possesses three properties: the function name, a list containing all 
the function’s parameter names and their corresponding types, and the return type.
After that, \Tool{} instructs the ERC to extract all event declarations from the ERC using a 
similar approach.

Second, \Tool{} iterates through each extracted function 
to extract rules for them individually. 
To prevent scenarios where a function contains numerous rules, 
and the LLM overlooks some when using only one prompt, 
\Tool{} employs multiple prompts for each function to extract different types of rules. 
As \Tool{} utilizes these extracted rules to analyze contract implementations, it selects implementation categories in Table~\ref{tab:linguistic}. 
Specifically, \Tool{} employs four prompts for each function, corresponding to the four pattern groups.
In each prompt, \Tool{} provides the entire ERC document, the function’s declaration, and a concise explanation of the linguistic group. Additionally, \Tool{} presents all identified patterns for the group as one-shot examples in the prompt. 
Moreover, \Tool{} instructs the LLM to present the extracted rule in a YAML format tailored to each group.
For the CP group, \Tool{} instructs the LLM to extract the condition, the condition type, and the action. For the EP group, 
\Tool{} directs the LLM to extract the condition and the event name. For the RP group, \Tool{} specifies that the LLM should extract the method for generating the return. Lastly, for the AP group, \Tool{} instructs the LLM to extract the action.

Take the rule violated in Figure~\ref{fig:20-high} as an example, 
the extracted condition is `` the \_from account has deliberately authorized the sender of the message via some mechanism'', the condition type is ``unless'', and the action is ``throw''.  

Third, \Tool{} examines each extracted event and extracts 
rules regarding when it should or should not be emitted, 
using a prompt similar to those designed for extracting rules in a linguistic pattern within the EP group from a function.

Following this process, \Tool{} successfully extracts 212 rules out of the 222 rules, misses 10 
rules, and mistakenly extracts four extra rules. 
%One missed rule is required by ERC20 to inspect 
%whether the returned Boolean value is \texttt{false} and take proper actions if so. This rule is 
%not associated with a function or an event and thus is not extracted by \Tool{}. For other 
%missed rules, we suspect they are due to the long ERC document provided in the prompt. We leave 
%the removal of unrelated parts of an ERC document to improve the effectiveness of rule 
%extraction as a future work.
%The four mistakenly extracted rules are caused by the same reason.
%Taking the one from ERC20 as an example, \Tool{} extracts a rule that \texttt{transferFrom()} should throw if the message caller’s account balance does not have enough tokens to spend. 
%ERC20 does not have such a requirement for \texttt{transferFrom()}, while it has this requirement for \texttt{transfer()}. We suspect \Tool{} makes this mistake because the two functions’ names are too similar. 
%The remaining three mistakenly extracted rules are from ERC3525.
%They are all caused by similar reasons.
%
We manually correct the errors that occurred during rule extraction to set up \Tool{} for 
subsequent violation detection. We emphasize the irreplaceable value of automated rule 
extraction due to two main reasons: 1) The LLM efficiently processes and condenses ERC documents 
into a concise format, and manual efforts are inherently constrained. 2) Extracted rules are 
reusable across all smart contracts implementing the ERC, reducing manual efforts to a one-time 
occurrence.

\subsection{Code Slicing}
A contract file might be excessively large, 
surpassing the input token limit of the LLM, 
making it impractical to input the entire file in a single prompt. 
Even if a contract file can fit within a prompt, supplying an extensive amount of code to 
the LLM could result in violations being buried in the input code, complicating the identification of the violations. Additionally, there is a risk that even for detected violations, the LLM may not report them due to the word limit of output. 
Consequently, 
\Tool{} divides each input contract file into smaller 
pieces and analyzes them individually.

We separate each contract 
based on its \texttt{public} functions as defined in the corresponding ERC 
for two reasons. First, after deployment, 
interactions with the contract occur through its \texttt{public} functions, and rule violations within 
the contract are exploited through those functions. 
Second, our study finds that, for the majority of rules, their scopes are confined to a \texttt{public} function (Insight 2), and even for those rules whose scopes are the 
entire contract, they can be verified 
by examining all \texttt{public} functions.

For each \texttt{public} function, \Tool{} calculates its associated code and 
mandates the LLM to analyze the related code together with the function. This approach is necessary because understanding the semantics of a function may be 
impractical solely by reading its code. We define a function's direct or 
indirect callees as its related code. Additionally, we consider contract 
fields accessed by the function or any of its callees as part of the related 
code. To distinguish functions defined in different contracts, we include the 
contract declaration of the \texttt{public} function or any of its callees. 
Lastly, given the LLM's proficiency in understanding natural languages, we 
incorporate comments associated with the function and its callees, both 
preceding and inside the function and its callees.

Taking the contract in Figure~\ref{fig:20-high} as an example, 
the entire contract contains 127 lines of code in total. 
We conduct a separate analysis of its nine \texttt{public} functions. 
When analyzing \texttt{transferFrom()}, we consider \texttt{\_transfer()} 
as its related code since it is called by \texttt{transferFrom()}. 
We also include line 2 as related code, 
as this field is accessed by \texttt{\_transfer()} in lines 14 and 15. 
Conversely, we do not consider line 3 as related code since the field is 
not accessed by \texttt{transferFrom()} and its callee. 
Furthermore, we incorporate the contract declarations in lines 1 and 18. 
Ultimately, the code analyzed by \Tool{} for \texttt{transferFrom()} 
contains 26 lines.

%\shihao{Specifically for interface-related auditing,  implementation code can be optimized out but only leave the interfaces as the context.}

\subsection{Optimization Mechanisms}

We have developed three mechanisms to enhance the LLM's understanding of ERC rules and rule violations. 
%by designing specialized 
%questions for each type of ERC rules, 
%breaking down compound rules, and providing 
%examples for rule violations.

\italicparagraph{Prompt Specialization}
Rather than issuing general inquiries to the LLM to determine if the input code violates a 
given rule, we craft specific questions 
tailored to the information stored in YAML format for each rule. 
This approach enhances the LLM's understanding of the rule, 
thereby improving effectiveness. For example, utilizing the information extracted for the violated rule in Figure~\ref{fig:20-high}, 
we instruct the LLM to examine whether each transferFrom() function 
throws unless "the \_from account has deliberately authorized the sender of the message via some mechanism."

\italicparagraph{One shot.}
To enhance the LLM's understanding of certain 
rule descriptions and Solidity program semantics, 
we employ one-shot learning. Particularly we incorporate a single example 
in the prompts for 36 rules. These examples are specifically designed for the rules 
and remain the same across various contracts.

The rules with a one-shot example can be categorized 
into three scenarios. First, we employ an example to elucidate a specific Solidity 
language feature for 30 rules. For instance, post version 0.5, 
Solidity introduced keyword \texttt{emit} for event emission. Prior to 
version 0.5, events were emitted akin to function calls. Consequently, we 
demonstrate how to emit events before version 0.5 
for rules necessitating event emission. %when checking contracts
%implemented in a Solidity version prior version 0.5.
Second, we utilize examples to guide the LLM's understanding of
implementing specific semantics for five rules. 
For example, ERC20 mandates the emission of a \texttt{Transfer} event for 
``initial token distribution.'' We include an example of an ERC20 contractor 
constructor to aid the LLM's comprehension. Additionally, we devise examples 
for rules requiring a function to return \texttt{false} when it fails to 
complete 
its functionality, preventing the LLM from erroneously equating throwing an exception with returning \texttt{false}.
Third, 
the remaining rule contains obscure descriptions. 
ERC20 stipulates that \texttt{transferFrom()} should verify whether 
the message sender 
has been authorized ``via some mechanism.'' 
We construct an example to illustrate that this mechanism should utilize the 
\texttt{\_allowances} field, like line 10 of Figure~\ref{fig:20-high}.

\italicparagraph{Breaking down compound rules.}
There are 24 rules structured in a way that dictates an action should be taken 
when a condition is met. 
One such rule is mandated by ERC20, 
specifying that a caller must verify the return value is equal to 
\texttt{false}, when a function returns a Boolean value. The remaining rules 
dictate that when a certain condition is met, 
a specific event must be emitted.

When directly querying the LLM about whether a function 
violates one of those compound rules, 
the LLM often interprets the absence of the 
prerequisite as a rule violation, resulting in a false positive. To 
address this, we break down each compound rule into two prompts.
The former prompt prompts the LLM to assess the existence of the condition. 
If the LLM confirms its presence, then \Tool{} sends to latter prompt to 
instruct the LLM to examine 
whether the action exists. \Tool{} reports a rule violation when the LLM 
considers the action is absent.

\section{Evaluation}
\label{sec:eva}

%\subsection{Methodology}

\bolditalicparagraph{Implementation.}
We employ GPT-4 Turbo~\cite{gpt4turbo} as the LLM in \Tool{} 
and interact with it using OpenAI's APIs. All other functionalities are 
implemented in Python, covering tasks such as automatically 
creating prompts, sending prompts to the LLM, 
and parsing the LLM’s responses. 
Notably, when constructing prompts, we require the LLM to inspect functions individually, 
rather than analyzing the entire Solidity contract. 
To achieve this, for each function, we perform 
static program analysis to compute both direct and 
indirect callees, contract fields referenced by the 
function or any of its callees, and contracts 
declaring the function and its callees. The results 
of the static analysis are then mapped back to the 
Solidity source code with the line-number 
information. Relevant source code lines and comments 
are extracted to form the prompts. 
The static analysis is conducted using the slither framework~\cite{slither}. 
%In total, \Tool{} comprises 2788 lines of source code, \shihao{
%    including 2140 lines of Python scripts and 648 lines of YAML files. Python scripts include smart contract static analysis powered by Slither, prompt generation, prompt executor, and result parser. YAML files are prompt templates.
%}

\bolditalicparagraph{Benchmarks.}
We construct two datasets for the evaluation.
The first one is a \emph{large} dataset with 200 contracts in total, including 
100 ERC20 contracts, 50 ERC721 contracts, and 50 ERC1155 contracts. 
We randomly collect ERC20 contracts from 
etherscan.io~\cite{etherscan} and ERC721 and ERC1155 
contracts from polygonscan.com~\cite{polygonscan}.
These two platforms are the most popular analytics 
platforms for Ethereum and its sidechain Polygon~\cite{polygon}, respectively.
In the contract sampling process, we specifically target contracts where both the contract itself and its associated Solidity code (\eg, libraries, inherited contracts) 
are contained in a single file to simplify the following experiments. 
On average, each contract file contains $847.7$ lines of Solidity source 
code.
Since the number of contract files is large, 
we do not manually analyze these contracts. 
We only inspect the results after applying \Tool{} on them.

The second dataset is a \emph{ground-truth} dataset with 30 ERC20 contracts. 
All these contracts undergo manual auditing by the Ethereum Commonwealth 
Security Department, a process where Solidity programmers submit auditing 
requests through filing an issue on GitHub, 
and the department subsequently provides auditing results by responding
to the issue~\cite{humanaudited}.
To form this dataset, we select the most recent 30 auditing requests 
meeting the following criteria: 1) providing the Solidity source code, 
2) approved by the Solidity programmers indicated by the ``approved’’ tag, 
3) containing ERC rule violations, 
and 4) having the contract and the Solidity code used 
by the contract in the same contract file. 
%Contract files associated with the same request 
%are merged into one Solidity file 
%to create the dataset. 
On average, each contract file contains 260.9 lines of 
Solidity source code. 
We carefully inspect these contracts and 
identify 142 ERC rule violations.
Among them, 21 violations have a high-security impact, 60 have a medium-security 
impact, and 61 have a low-security impact.

\bolditalicparagraph{Baseline Techniques.}
We choose two baselines.
The first one is the auditing service provided by Ethereum Commonwealth 
Security Department (ECSD). This is the only service 
whose auditing results are available to us.
We compare \Tool{}’s results with the auditing results provided by
their Solidity security experts.
%\yiying{explain why choosing this auditing service, why it is representative of all auditing services (and thus results can generalize)}
The second baseline is slither-check-erc (SCE).
As discussed in Section~\ref{sec:related}, it is the most powerful open-source technique to validate ERC conformance, 
and it covers all ERCs involved in our benchmarks. 
%\yiying{similarly, explain why choosing this tool, why it is representative of all tools (and thus results can generalize)}

\bolditalicparagraph{Research Questions.}
Our experiments are designed to answer the following research questions:
%
%\begin{itemize}
%
1) \emph{Effectiveness}: can \Tool{} accurately pinpoint ERC rule violations? 
2) \emph{Advancement}: does \Tool{} perform better than existing auditing solutions?
3) \emph{Necessity}: how does each component of \Tool{} contribute to its auditing capability?

%\end{itemize}

\bolditalicparagraph{Experimental Setting.}
We configure the temperature parameter to zero to make our experimental results
stable when interacting with the LLM. 
All our experiments are performed on a desktop machine, with Intel(R) Core(TM) i7-10700 CPU, 64GB RAM, and Ubuntu 22.04 OS version.

\subsection{Effectiveness of \Tool{}}

\begin{table}[t]
\centering
\small

\mycaption{tab:large}
{Evaluation results on the large dataset.}
{\textit{($(x, y)$: $x$ true positives, and $y$ false positives.)}
}
{
\begin{tabular}{|l|c|c|c||c|}
\hline
                 &  {\textbf{High}} & {\textbf{Medium}} & {\textbf{Low}}  & {\textbf{Total}} \\ 

%\cline{2-3}
%\cline{4-5}
%\cline{6-7}
%\cline{8-9}

% & {{\textbf{TP}}} &  {{\textbf{FP}}}  & {{\textbf{TP}}} &  {{\textbf{FP}}}  & {{\textbf{TP}}} &  {{\textbf{FP}}} & {{\textbf{TP}}} &  {{\textbf{FP}}} \\

\hline
\hline

{\textbf{ERC20}}     & $(1, 0)$ & $(97, 1)$  & $(29, 8)$    & $(127, 9)$    \\ \hline
{\textbf{ERC721}}    & $(0, 1)$ & $(3, 3)$   & $(109, 1)$   & $(112, 5)$  \\ \hline
{\textbf{ERC1155}}   & $(3, 0)$ & $(12, 0)$  & $(25, 1)$    & $(40, 1)$ \\ \hline \hline
{\textbf{Total}}     & $(4, 1)$ & $(112, 4)$  & $(163, 10)$  & $(279, 15)$ \\ \hline

\end{tabular}
%}
}

% \vspace{-0.2in}
\end{table}

\bolditalicparagraph{Methodology.}
We execute \Tool{} on the large dataset and count violations and false positives reported by \Tool{} to gauge its effectiveness. 
For each violation flagged by \Tool{}, we manually examine the description generated by the LLM and the corresponding smart-contract code to 
determine whether \Tool{} accurately identifies a true violation 
or if it reports a false alarm. Each reported violation is analyzed by at least two paper authors. Any disagreements are 
resolved through multiple rounds of discussion. 
As the total number of violations in the large dataset is unknown to us, we do not assess false negatives in this experiment.

{
\begin{figure}[t]

\begin{minipage}{\columnwidth}
\begin{center}
\scriptsize
\begin{lstlisting}[numbers=left,framexleftmargin=.15in,xleftmargin=.15in,language=Solidity,basicstyle=\ttfamily,morekeywords={-},morekeywords={+},keepspaces=true]
 contract MyERC1155Token {
   mapping(uint256 => mapping(address => uint256)) _balances;
   mapping(address => mapping(address => bool)) _opApprovals;

   function safeTransferFrom(address from, address to, uint256 id, uint256 value, bytes memory data) public {
+    require(_opApprovals[from][msg.sender], "the caller is not approved to manage the tokens");
     _update(from, to, id, value);
+    if (to.isContract()) {
+        try IERC1155Receiver(to).onERC1155Received(msg.sender, from, id, amount, data) returns (bytes4 response) { ... }
+    }
   }
   function _update(address from, address to, uint256 id, uint256 value) internal {
     if (from != address(0)) {
       _balances[id][from] -= value;
     }
     if (to != address(0)) {
       _balances[id][to] += value;
     }
   }
 }
\end{lstlisting}
\vspace{-0.05in}
\mycaption{fig:1155-high}{An ERC1155 rule violation with a high-security
impact.}
{\textit{(Code simplified for illustration purpose.)}}
\end{center}
\end{minipage}
\vspace{-0.25in}
\end{figure}
}

\bolditalicparagraph{Effectiveness Results.}
As shown in Table~\ref{tab:large}, 
\Tool{} detects 279 ERC rule violations, including four with a high-security impact,
112 with a medium-security impact, 
and 163 with a 
low-security impact. Additionally, \Tool{} reports 15 false positives. 

\Tool{} identifies four violations with a high-security impact. 
One of the violations is pinpointed from an ERC20 contract 
as shown in Figure~\ref{fig:20-high},
while the other three are associated with two ERC1155 contracts. 
Figure~\ref{fig:1155-high} illustrates one of the ERC1155 contracts. 
The function \texttt{safeTransferFrom()} in lines 5–8 is designed to 
transfer \texttt{value} amounts of tokens of a specific type 
(indicated by parameter \texttt{id}) from one address to another. This process involves decreasing 
the balance of \texttt{from} in line 11 and increasing the 
balance of \texttt{to} in line 14. ERC1155 mandates that the caller 
of \texttt{safeTransferFrom()} must be approved to manage the tokens 
for the token owner. Unfortunately, neither \texttt{safeTransferFrom()} 
nor its callee \texttt{\_update()} performs this crucial 
check (via \texttt{\_opApprovals}), allowing anyone to transfer 
tokens from any address to their own.
The patch in line 6 introduces a verification step to ensure that 
the token owner \texttt{from} has approved the message 
sender to handle his tokens, thus aligning with the ERC1155 requirement.
Moreover, ERC1155 mandates that \texttt{safeTransferFrom()} checks whether 
the recipient is a contract and if so, whether the contract is capable of handling
ERC1155 tokens.
Unfortunately, the implementation in Figure~\ref{fig:1155-high} neglects these
checks. Without the required capability, transferring tokens to a recipient contract 
results in the tokens being indefinitely trapped within the contract.
Lines 8-10 show the necessary checks to fulfill ERC1155 requirements. 
The final high-security impact violation arises from the failure to check 
whether the recipient of a token transfer operation is address zero. 
Transferring tokens to address zero leads to the irreversible loss of the 
tokens.
Overall, \emph{\Tool{} is capable of detecting violations of ERC rules that can result in the loss of digital assets. }

The 112 violations with a medium-security impact stem from various reasons. 
Among them, 82 violate an ERC20 rule that dictates the function 
\texttt{transfer()} and \texttt{transferFrom()} should treat the transfer of zero values as a normal 
transfer. However, in 83 cases, 
\texttt{transfer()} or  \texttt{transferFrom()}
of an ERC20 contract implementation throws an exception when the message caller attempts to send zero 
tokens. 14 violations occur due to the absence of implementation for a 
function mandated by the corresponding ERC. Nine violations involve a function 
lacking the required return-value type specified by the ERC or having no 
return at all, in contrast to ERC requirements. Three violations result from 
neglecting to check the return of a function that returns a Boolean value, as 
mandated by ERC20.
An additional three violations arise from the failure to throw an exception 
while implementing the \texttt{ownerOf()} function of ERC721. The remaining 
violation is also depicted in Figure~\ref{fig:1155-high}, as ERC1155 requires 
\texttt{safeTransferFrom()} to utilize its input parameter \texttt{data} to 
call \texttt{onERC1155Received()}.
These findings underscore \emph{\Tool{}'s capability to identify various types of violations.}

Of the 155 event-related violations, 146 fail to emit an event, and 9 are due to the omission of an event declaration. 

In general, \emph{\Tool{} exhibits accuracy}, 
with only 15 reported false positives for four reasons.
First, in three cases, the input code 
is too long (exceeding 200 lines) or 
too complex (using a variable in 
multiple function calls and condition 
checks). Consequently, \Tool{} mistakenly concludes that a required action is not taken. 
Second, three false positives arise 
from \Tool{} failing to understand the \texttt{require} statement 
triggers an exception when its input 
condition is unsatisfied. 
%which causes 
%3 false positives, and \Tool{} does not know subtraction induces an exception in the event of underflow, leading to 2 false positives. \mengting{no fp about subtraction now}
Third, \Tool{} makes errors when 
inferring program semantics, resulting 
in eight false positives. For example, 
for seven false positives,  the 
inspected code invokes an external 
function named \texttt{transfer()}, 
not necessarily implying the transfer 
of tokens. Despite this, \Tool{} 
incorrectly insists on the need for the code snippet to emit a transfer event.
Fourth, in the remaining case, the input 
function unconditionally triggers an 
exception, which, in our assessment, 
satisfies the rule since an exception 
is indeed triggered when the condition 
is met. However, \Tool{} reaches a different conclusion.

%\shihao {
%We exclude the 81 results in ERC1155 smart contracts from three rules due to one reason: their requirements do not stand in certain valid scenarios. For example, ERC1155 requires the `\_from` argument MUST be the address of the holder whose balance is decreased for the event TransferSingle. However, this is incorrect in the minting operation. The from address is usually zero and its balance does not decrease. Similarly, the `\_to` argument is not the address of the recipient whose balance is increased when in a token-burning operation, which is another example of a problematic rule. 
%}

\begin{table}[t]
\centering
\small

\mycaption{tab:compare}
{Evaluation results on the ground-true dataset.}
{\textit{($(x, y, z)$: x true positives, y false positives, and
z false negatives, and Columns Time and Money represent 
the total time and monetary cost.)}
}
{
\setlength{\tabcolsep}{0.12mm}{
\begin{tabular}{|l|c|c|c||c|c|c|}
\hline
                 &  {\textbf{High}} & {\textbf{Medium}} & {\textbf{Low}} &
                 {{\textbf{Total}}} &
                 {{\textbf{Time(s)}}} & {{\textbf{Money}}} \\ 

%\cline{2-3}
%\cline{4-5}
%\cline{6-7}

% & {{\textbf{$TP_{FP}$}}} & {{\textbf{FN}}} & {{\textbf{$TP_{FP}$}}} & {{\textbf{FN}}} & {{\textbf{$TP_{FP}$}}} & {{\textbf{FN}}} & & \\

\hline
\hline

{\textbf{ECSD}}  & $(10, 10, 11)$ & $(26, 2, 34)$ & $(37, 0, 24)$ & $(73, 12, 69)$ & $2.6*10^7$  & \$$1.5*10^5$  \\ \hline
{\textbf{SCE}}   & $(0, 0, 21)$ & $(27, 0, 33)$ & $(12, 0, 49)$  & $(39, 0, 103)$ & 0.02298  & -   \\ \hline
{\textbf{\Tool{}}}   & $(21, 1, 0)$  & $(57, 1, 3)$  & $(61, 1, 0)$ & $(139, 3, 3)$ & 1780.1 & \$11.92   \\ \hline

\end{tabular}
}
}

% \vspace{-0.2in}
\end{table}

\subsection{Comparison with Baselines}
\label{sec:compare}

\bolditalicparagraph{Methodology.}
We perform a head-to-head comparison between \Tool{}
and the two baseline solutions using the ground-truth dataset.
We execute \Tool{} and SCE on the ground-truth dataset.
We manually examine their results and the auditing reports 
from ECSD security experts
to count true positives, false positives, 
and false negatives for each tool.
We assess the solutions’ effectiveness and accuracy based on these statistical measures. 

Moreover, we assess the costs associated with each solution in terms of time and money. 
For \Tool{} and SCE, we execute each tool on every contract file three times and report the average execution time. 
The time spent by ECSD is measured from the moment a Solidity programmer submits an auditing request by filing an issue report on GitHub until a security expert provides the auditing result by responding to the issue, considering that the Solidity programmer must wait for this duration to receive the result. 
The monetary cost of \Tool{} is represented by the fees charged by OpenAI. 
We only have expense information for one auditing request in the dataset, which amounts to \$1000. 
We calculate the average expense per hour by dividing 1000 by the number of hours security experts spent on the contract.
%Thus, we randomly sample another nine auditing requests with expense information (ten in total). 
%Subsequently, we compute the average expense per hour. 
For each contract file without expense information, we multiply the average expense per hour with 
the time spent on the auditing request to estimate the monetary cost.
Since SCE is an open-source software, 
no monetary expenditure is associated with its use. 

\bolditalicparagraph{Effectiveness and Accuracy.}
As shown in Table~\ref{tab:compare},
\Tool{} detects nearly all violations with just three false negatives and reports only 
three false positives. 
Compared with the two baseline solutions, 
\emph{\Tool{} demonstrates greater effectiveness by 
identifying more true violations and increased accuracy with fewer false positives. }

\italicparagraph{\Tool{}} accurately identifies all 21 violations with a high-security impact. 
Among them, nine stem from a failure to verify whether the message sender possesses the required privilege, as mandated by ERC20,
mirroring the violation in Figure~\ref{fig:20-high}. 
%\mengting{One results from incorrectly updating the value of allowance in function approve. It increases the value of allowance by '\_value', which should be directly assigned with '\_value'.}
One ERC20 rule requires function \texttt{approve(address \_spender, uint256 \_value)} to overwrite the allowance value of \texttt{\_spender}
with the input \texttt{\_value}.  
\Tool{} pinpoints a case where \texttt{approve(address \_spender, uint256 \_value)}
increases the allowance of \texttt{\_spender} by \texttt{\_value}, thus violating the rule. 
The remaining 11 violations result from neglecting 
to check if the sender holds a sufficient balance 
when transferring tokens. Notably, Solidity 
introduced the underflow check for subtraction 
operations in version 0.8.0. Consequently, Solidity 
programmers must explicitly compare the balance with 
the transferred amount for versions predating 0.8.0, 
and failing to do so leads to the 11 identified 
violations.
\Tool{} also uncovers 13 instances with a medium-security impact where 
a function fails to generate its return as specified in ERC20. 
For instance, ERC20 dictates that \texttt{totalSupply()} 
should return the quantity of supplied tokens. 
However, in one violation, the implementation of \texttt{totalSupply()} 
returns the result of subtracting the balance of address zero from the supplied tokens. 
For all other violations with a medium- or low-security impact, 
\Tool{} identifies similar issues within the large dataset.

\Tool{} misses three violations. Two are in the same 
the implementation of \texttt{transfer()}, which conducts an overflow check. 
This check uses a \texttt{require} statement to 
mandate the sum of the receipt's balance and the 
token amount must be greater than the receipt's 
balance, resulting in the inability to transfer zero 
tokens, thus violating an ERC20 rule. Unfortunately, 
\Tool{} fails to comprehend such program semantics. 
The remaining false negative is attributed to the 
excessive length of the input code, consisting of 
180 lines. This length causes \Tool{} to miss 
checking the returned Boolean value of an external 
function call.

\Tool{} reports three false positives with reasons 
distinct from those in the large dataset. 
Two of the false positives emerge in a contract's \texttt{transferFrom()} 
function. This function should subtract the token amount from the current 
value of an element in the allowance map and utilize the result to update the element (like line 
5 in Figure~\ref{fig:20-high}). However, the implementation contains a bug 
wherein it subtracts the token amount from the value of an element in the balance map and uses the result 
to update the element in the allowance map. 
We postulate that these two false positives 
result from this bug, as \Tool{} does not recognize that the \texttt{transferFrom()} implementation checks whether the message sender has the privilege using the allowance map. Additionally, it considers the 
function’s return does not match whether it completes the transfer.
Regarding the remaining false positive, we cannot ascertain the reason. The 
code is relatively straightforward, and the \texttt{approve()} function emits 
the required \texttt{Approval} event. Nonetheless, \Tool{} reports that the 
necessary event is not emitted.

\italicparagraph{ECSD.}
The manual auditing service identifies 73 violations. 
Only two violations, where an overflow check causes the inability to handle transferring zero tokens, 
are missed by \Tool{}. Additionally, 
\Tool{} identifies 68 violations that the auditing service cannot detect. 
Furthermore, the service has 12 false positives, 
exceeding the number in \Tool{}. Two false positives occur because auditors are unaware that Solidity automatically generates getter functions for contract fields, leading them to mistakenly believe that two required APIs are not implemented. ERC20 mandates \texttt{transfer()} to check if the token owner 
has enough balance for the transfer, but no such rule applies to \texttt{transferFrom()}. However, security auditors mistakenly assume a similar rule exists for \texttt{transferFrom()}, resulting in the remaining ten false positives. In conclusion, \emph{\Tool{} demonstrates superior violation detection capability and accuracy compared to the manual auditing service.}

\italicparagraph{SCE.}
As explained in Section~\ref{tab:compare}, 
the functionality of SCE is relatively 
straightforward, leading it to overlook the most 
violations among the three solutions. All 39 violations detected by SCE are also identified by 
\Tool{}. Among these violations, 27 involve 
instances where a function mandated by ERC20 is 
either not implemented or does not match the API 
specified in ERC20. The remaining 12 violations are 
event-related. Given that the rules covered by SCE 
are uncomplicated, 
SCE does not generate any false positives.

\bolditalicparagraph{Time and Monetary Cost.}
Among the three solutions, SCE incurs the smallest cost. 
It analyzes the 30 contract files in the ground-truth dataset in less than 0.1 
seconds and does not involve any charges. 
However, due to its limitation in detection capability, 
SCE is inadequate in identifying ERC rule violations.

\Tool{} requires interaction with the LLM provided by OpenAI, resulting in a longer processing time compared to SCE, 
which solely conducts static program analysis. 
Conversely, \Tool{} incurs significantly lower costs compared to the manual service provided by ECSD. 
The time usage of the manual service is $14535$ times higher, 
and the monetary usage is $12562$ times greater. \emph{\Tool{} outperforms the manual service with significantly lower costs.}

%\mengting{Slither-check-erc could check the erc's conformance. We ran this tool on the small dataset. It reported 40 inconsistencies. All of them were real inconsistencies. 12 of them were mismatch of return types of optional functions such as decimals. For example, DSToken defines decimals in uint256, but it should be uint8. 9 of them were functions like transfer not returning bool. 6 of them were missing functions. 7 of them were missing event. 4 of them were not emitting required event in functions. The other 2 were parameters of event Transfer were not indexed. All these can be detected by our tool.}

% \input{figures/fig-fp}
\vspace{-0.1in}

\subsection{Necessity of \Tool{} design points}

\bolditalicparagraph{Methodology.}
To comprehend the significance of individual design points, 
namely ERC rule sequentialization, prompt specialization, 
code slicing, one-shot, and breaking down compound rules, 
we utilize \Tool{} on the ground-truth dataset with each design point deactivated. 
We then count the number of true positives and false positives 
in each experimental setting.
%\songlh{XXX}

\begin{figure}[t]
\centering
\includegraphics[width=0.8\columnwidth]{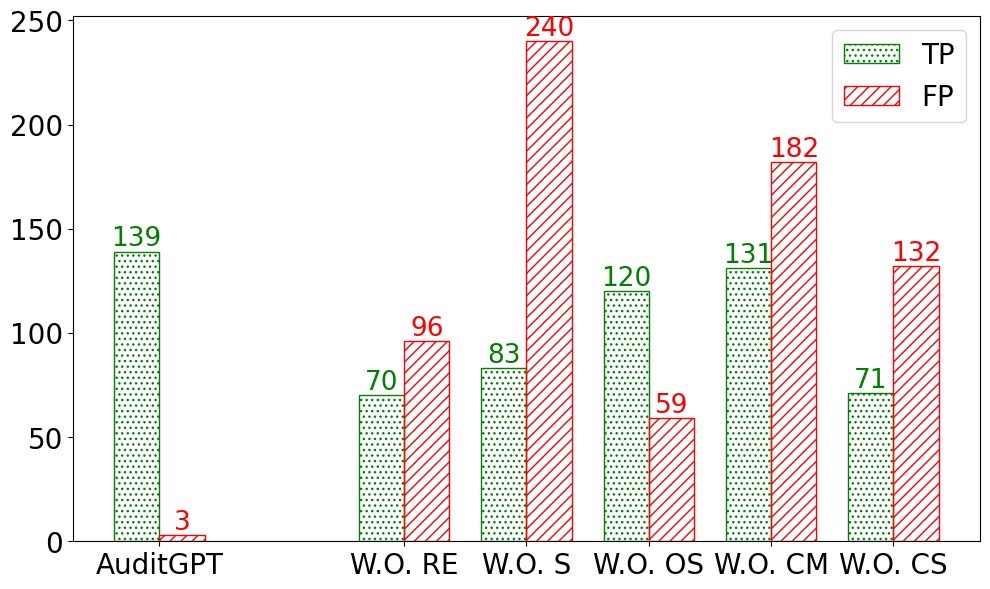}
\mycaption{fig:tp-fp-bar}{Evaluation results on the ground-truth dataset with each design point deactivated.}
{\textit{(W.O.: without, RE: rule sequentialization, S: specialization, OS: one shot, CM: breaking down, CS: code slicing)}
}

\vspace{-0.1in}
\end{figure}

\bolditalicparagraph{Experimental Results.}
As shown in Figure~\ref{fig:tp-fp-bar}, the full-featured \Tool{} identifies
the highest number of violations and reports the fewest false positives, 
thereby showcasing \emph{the rationale behind each design point of \Tool{}.} 
Additionally, the full-featured version detects all violations identified by other versions, 
except for one. This particular violation stems from the failure 
to check the returned Boolean value of an external call. 
While the full-featured version misses it, the version with disabled code 
slicing captures it, indicating that a more advanced setting does not necessarily yield better results in all cases.

Both ERC rule sequentialization transforms the process of examining 
the entire ERC20 document (with 35 rules in total) 
into scrutinizing individual rules. 
Similarly, code slicing breaks down the entire contract 
into individual functions. Both approaches simplify the auditing problem 
within a single prompt. Consequently, disabling either of them 
would lead to \Tool{} missing the most violations among all settings. 
Prompt specialization reduces the difficulty for the LLM to understand a ERC rule 
and aids the LLM in comprehending when the rule is violated. 
Without it, \Tool{} reports the largest number of false positives among all settings.
Five ERC20 rules require a specific action when a condition is 
met. Without breaking down compound rules, \Tool{} considers many cases where 
a condition does not exist as rule violations, leading to 182 false positives.
One shot assists the LLM in understanding what constitutes a violation for a particular rule with a concrete example. It helps reduce 19 false negatives and 56 false positives.

\subsection{Threats to Validity} 
Threats to validity arise from two primary sources. First, we randomly sample smart contracts 
for our experiments, and we don’t particularly include contracts 
with intricate control flow or data dependence relations. We anticipate LLMs might exhibit worse performance when auditing code with such complexities. 
Second, our empirical study is confined to four ERCs, 
and our experiments only encompass three of them. 
Consequently, the findings and observations from our study and experiments 
may not be generalized to other ERCs. 
Evaluating \Tool{} with more intricate code and 
additional ERCs remains a topic for future investigation.

%\mengting{
%The main internal threat to validity comes from the complexity of the code. ChatGPT does not understand complex code well. If the code is too long or has a complex structure, ChatGPT has a higher possibility of giving a wrong answer. Another internal threat comes from the quality of ERC, specifically,  standardization and accuracy. \\
%The main external threat comes from the lack of diversity in ERC. AuditGPT only audits ERC-20, ERC-721, and ERC-1155 tokens. It may not generalize to other ERC tokens.\\
%It can also be extended to check code compliance and policy compliance. Extract rules from code specifications of other programming languages and ask ChatGPT to analyze code and identify violations. Extract rules from policies in other areas such as finance and health and ask ChatGPT to analyze records or reports and identify issues. 
%}

\section{Conclusion}

%Facing the increasing popularity of blockchain systems and smart 
%contracts, 
In this paper,
we conduct an empirical study on the implementation 
rules outlined in ERCs, considering four key aspects: 
their content, security impacts, specifications in ERCs, 
and potential implementations in Solidity programs. 
Using the study insights, we develop an 
automated tool called \Tool{} that employs LLMs to compare smart 
contracts with their corresponding ERCs and identify rule violations.
\Tool{} successfully pinpointed numerous rule violations, 
outperforming the two baseline solutions. We anticipate that this 
research will enhance the understanding of ERC rules and their 
violations, inspiring further exploration in this 
field. Future work may investigate leveraging LLMs to determine 
whether a code implementation aligns with a natural language 
description in other scenarios, as well as detecting other types 
of Solidity bugs.

\bibliography{gpt-1}
\bibliographystyle{abbrv}
\end{document}